\documentclass[10pt,letterpaper,aps,prl,reprint,superscriptaddress,showpacs,twocolumn,nofootinbib,floatfix,amsmath,amssymb]{revtex4-1}

\usepackage{amsmath}
\usepackage{amsfonts}
\usepackage{amssymb}
\usepackage{mathrsfs}

\usepackage{color}
\usepackage{hyperref}

\DeclareMathOperator{\tr}{tr}

\def\B{\mathcal{B}}
\def\D{\mathcal{D}}

\begin{document}

\title{Entanglement Entropy of Electromagnetic Edge Modes}

\author{William Donnelly}
\affiliation{
Department of Physics,
University of California Santa Barbara,
Santa Barbara, California 93106, USA}
\email{williamdonnelly@gmail.com}
\author{Aron C. Wall}
\affiliation{
School of Natural Sciences,
Institute for Advanced Study,
Princeton, New Jersey 08540, USA}
\email{aroncwall@gmail.com}

\begin{abstract}
The vacuum entanglement entropy of Maxwell theory, when evaluated by standard methods, contains an unexpected term with no known statistical interpretation. We resolve this two-decades old puzzle by showing that this term is the entanglement entropy of edge modes: classical solutions determined by the electric field normal to the entangling surface. We explain how the heat kernel regularization applied to this term leads to the negative divergent expression found by Kabat. This calculation also resolves a recent puzzle concerning the logarithmic divergences of gauge fields in 3+1 dimensions.
\end{abstract}

\pacs{
04.70.Dy 
11.15.-q 
}

\maketitle

\paragraph{Introduction}
Entanglement entropy is relevant to black hole thermodynamics \cite{Sorkin1983, Bombelli1986, Srednicki1993, Wall2011, Solodukhin2011, Harlow2014}, 
quantum gravity \cite{Bianchi2012, Bousso2014},
gauge-gravity duality \cite{Ryu2006a, Hubeny2007, Nishioka2009, VanRaamsdonk2010, Faulkner2013, Swingle2014, Engelhardt2014},
field theory \cite{Casini2008,Casini2009,Blanco2013},
condensed matter physics \cite{Vidal2002,Kitaev2005,Levin2005,Hastings2007,Eisert2008,Grover2013},
$c$ theorems \cite{Casini2004,Casini2006c,Casini2012,Grover2012,Solodukhin2013},
and confinement \cite{Nishioka2006,Klebanov2007,Lewkowycz2012}.
But there are subtleties that arise for gauge fields that are not present in the case of scalar and spinor fields.
In $D$ spacetime dimensions, standard Euclidean methods give an area-law entropy divergence equal to that of $(D-2)$ scalar fields, 
plus a ``contact term'' coming from an interaction with the entangling surface, which until now has had no known statistical interpretation \cite{Kabat1995}.  This contact term may or may not appear depending on the method of calculation, which has led to a great deal of controversy about its significance  \cite{Barvinsky1995,DeNardo1996,Iellici1996,Cognola1997,Kabat2012,Donnelly2012,Solodukhin2012}.  If one includes the contact term, the divergences in the Euclidean entropy due to gauge fields are related to the renormalization of the gravitational action \cite{Larsen1995,Cooperman2013} in the manner expected from consistency of black hole thermodynamics \cite{Susskind1994, Jacobson1994, Fursaev1994, Demers1995, deAlwis1995,Frolov1996}.

When evaluated by heat kernel regularization as in \cite{Kabat1995}, the leading divergence of the contact term is negative.
This is disturbing, since entropy is a manifestly positive quantity.
However, different ways of calculating entropy can differ by local counterterms \cite{Casini2006a,Casini2006b,Casini2006c}, so the sign of the leading order divergence is not universal under a change of regulator scheme.
A more significant question is whether the geometric entropy is equal to the von Neumann entropy of the reduced density matrix, up to such local counterterms.
If this is not the case, it is unclear whether the geometric entropy should obey standard properties such as strong subadditivity \cite{Wehrl1978}.

One approach is to calculate entanglement entropy using a physical regulator, such as a lattice \cite{Srednicki1993}.
In a lattice scalar field theory we introduce Hilbert spaces $\{ \mathcal{H}_n : n \in N \}$ at each lattice site $n$.  
For any subset $A$ of lattice sites, we can restrict the total state $\rho$ to $A$ by tracing out degrees of freedom: $\rho_A = \tr_{\mathcal{H}_{N \setminus A}} \rho$.
The entanglement entropy of $A$ is the von Neumann entropy 
\begin{equation}
S(\rho_A) = - \tr \rho_A \ln \rho_A.
\end{equation}
In a lattice gauge theory, the Hilbert space is not a tensor product over regions of space, and there are many possible definitions of entanglement entropy \cite{Buividovich2008b,Donnelly2011,Casini2013,Donnelly2014,Casini2014}.
One natural definition embeds the Hilbert space into a tensor product of Hilbert spaces that include \emph{edge modes} living on the boundary \cite{Donnelly2011}.
These edge modes give a positive contribution to the entropy that depends only on boundary correlation functions.
It is expected that such a term will persist in the continuum limit, but to our knowledge, such a continuum limit has not previously been taken.

The issue of Hilbert space factorization is typically avoided by the use of Euclidean field theory methods.
For a planar entangling surface in flat space, the vacuum reduced density matrix has the thermal form $\rho \propto e^{-2 \pi K}$ with $K$ the boost generator\cite{Bisognano1975,Kabat1994,Jacobson1994b}.
The \emph{geometric entropy} is 
\cite{Callan1994}
\begin{equation} \label{conical}
S = \left. (1 - \beta \partial_\beta) \ln Z \right|_{\beta = 2 \pi},
\end{equation}
where $Z$ is the Euclidean partition function on a manifold with conical angle $\beta$.
This formula gives the correct answer for the black hole entropy \cite{Iyer1995}, but it is unclear whether it agrees with the von Neumann entropy.
The reason is that for $\beta \neq 2 \pi$ the manifold has a conical singularity, which can be regulated by smoothing \cite{Nelson1994,Fursaev1995,Frolov1995}. 
Kabat's contact term appears as a consequence of the gauge field coupling to this curvature \cite{Kabat1995,Larsen1995,Donnelly2012}.

We will regulate the conical singularity in a different way, by introducing a small regulating surface (a ``brick wall'' \cite{tHooft1985}) at a distance $\epsilon$ from the entangling surface.
This eliminates any possible coupling to the conical singularity and ensures that the geometric entropy \eqref{conical} coincides with the von Neumann entropy of the outside density matrix.
Standard boundary conditions at the brick wall suppress certain fluctuations of the field, changing physics far from the wall and underestimating the entanglement entropy.
Instead, we allow for arbitrary fluxes through the brick wall, leading to a sum over edge modes of the kind that appears on the lattice \cite{Donnelly2011}.
In the limit $\epsilon \to 0$ the brick wall entropy is equal to the geometric entropy \eqref{conical}, with the contact term included in the sum over edge modes.
Thus we find that the contact term has a statistical interpretation: it is the entanglement entropy of the edge modes.

We also discuss the implications for the logarithmic terms in the entanglement entropy.  We show that inclusion of the edge modes resolves the apparent discrepancy between the entanglement entropy of gauge fields and the corresponding conformal anomaly found by Dowker \cite{Dowker2010} (cf. \cite{Eling2013}).  We will also explain why the standard heat kernel regulator assigns a negative entropy value to the edge modes.

\paragraph{Edge modes}

Gauge theories are distinguished by the presence of constraints that restrict the space of physical states.
A state $\psi(E)$ in the electric field basis is supported on solutions of Gauss' law $\nabla \! \cdot \! E = 0$.
Gauss' law forces the perpendicular component of the electric field $E_\perp$ to match across the entangling surface, and this leads to a density matrix that commutes with $E_\perp$.
As a consequence, the reduced density matrix has a block-diagonal structure in the $E_\perp$ basis and the entropy is a sum of two terms \cite{Donnelly2011}:
\begin{equation} \label{entropy}
S = \int \D E_\perp\,p(E_\perp) \left[ - \ln p(E_\perp) + S(\rho_{E_\perp})\right]
\end{equation}
where $p_E$ is the classical probability distribution of $E_\perp$, and $\rho_{E_\perp}$ is the reduced density matrix in the sector with $E_\perp$ fixed.
Since the theory is free, $S(\rho_{E_\perp})$ is independent of $E_\perp$ and can be calculated with the boundary condition $E_\perp = 0$.
Note that the quantities appearing in \eqref{entropy} are formal and require a regulator to be well defined.

We now calculate the first term in \eqref{entropy}, which is the classical entropy associated to the edge modes in the vacuum state \cite{Balachandran1995}.
For each $E_\perp$, we find the unique classical solution of the form $E = \nabla \varphi$ satisfying the boundary condition $\nabla_\perp \varphi = E_\perp$.
In the vacuum, the edge modes are thermally populated according to their boost energy at temperature $(2 \pi)^{-1}$, so the entropy can be computed using \eqref{conical} from the partition function 
\begin{equation} \label{Zedge}
Z_\text{edge} = \int \D E_\perp e^{-I(E_\perp)},
\end{equation}
where $I$ is the on-shell Euclidean action.

We now specialize to the case of flat spacetime, although the result can easily be generalized to curved spacetimes \cite{Donnelly:2015hxa}. 
In order to simplify infrared physics, we compactify the $(D-2)$ transverse spacetime dimensions into a torus $\cal T$. 

The mode with constant $E_\perp$ has an infinite energy in flat spacetime, and hence vanishes exactly.  So it does not contribute to the entropy.

The nonconstant edge modes of $E_\perp$ have fluctuations that contribute to the entropy.
To regulate these fluctuations we introduce a brick wall at $r = \epsilon$, on which $E_\perp$ is fixed, and consider the leading order behavior as $\epsilon \to 0$.
As we take the limit $\epsilon \to 0$, the electric field extending from a given $E_\perp$ decays like $1/\ln(\epsilon^{-1})$ at fixed distance from the entangling surface.  Thus the boost energy becomes small, and the thermal fluctuations in each mode of $E_\perp$ become large.

In order to have well-defined dynamics at the brick wall, we need to choose a boundary condition. 
Neither of the two standard boundary conditions for Maxwell theory preserves physics far from the wall: an electrical conductor does not permit a magnetic flux and vice versa. 
Our solution is to impose the magnetic conductor boundary conditions $E_\perp = B_\parallel = 0$, and to explicitly sum over the missing $E_\perp$ modes using \eqref{Zedge}.

The metric in the normal directions takes the form
$ds^2 = dr^2 + r^2 d \tau^2$
where $\tau$ is periodic with period $\beta$.
We consider a mode expansion of $E_\perp$ and the vector potential $A$:
\begin{equation}
E_\perp = \sum_n E_n \psi_n(x), \quad A= \sum_n \phi_n(r) \psi_n(x) d \tau
\end{equation}
where $\psi_n(x)$ are scalar eigenmodes of the Laplacian $-\nabla^2_\mathcal{T}$ on the torus $\cal T$,
with eigenvalues $\lambda_n$.

As $r \to 0$ the solutions of the equation of motion $\nabla_a F^{ab} = 0$ behave as 
\begin{equation}
\phi_n(r) = \frac{E_n}{\ln(\epsilon^{-1})} \left( \frac{1}{\lambda_n} + \frac12 r^2 \ln r\right) + \mathcal{O}(r^2).
\end{equation}
The on-shell action of this solution is 
\begin{equation}\label{onshell}
I(E_\perp) = \frac12 \oint_{r = \epsilon} A \wedge \star F = \sum_n \frac{\beta E_n^2}{2 \lambda_n \ln(\epsilon^{-1})}.
\end{equation}

We now must integrate the on-shell action with the path integral measure $\D E_\perp$, which we define by taking the continuum limit of the discrete measure on the lattice.
On the lattice we have $N$ points $x$ to which we associate area $\mathrm{Area}(\mathcal{T})/N$.
In a $U(1)$ gauge theory with minimal charge $q$, the integrated flux through each area element is quantized in units of $q$.
The flux $E_\perp(x)$ at each point is therefore quantized in units of $qN / \mathrm{Area}(\mathcal{T})$; changing variables to the coefficients $E_n$ of the mode expansion yields the measure
\begin{equation} \label{measure}
\D E_\perp = \frac{1}{\sqrt{N}} \prod_{n > 0} \frac{1}{q} \sqrt{\frac{\mathrm{Area}(\mathcal{T})}{N}} d E_n,
\end{equation}
where we integrate over all modes except for the constant mode $n=0$.

We can now carry out the functional integral mode by mode.
Rescaling the determinant by $\zeta$-function regularization (up to a local anomaly in even dimensions) we find 
\begin{equation} \label{Zedge2}
Z_\text{edge} =
\det{}' \left( \frac{\ln (\epsilon^{-1})}{\beta} \frac{2 \pi \mathrm{Area}(\mathcal{T})}{q^2} (-\nabla^2_\mathcal{T})
\right)^{1/2}.
\end{equation}
The full entropy \eqref{entropy} is obtained from the full partition function, which is a product of $Z_\text{edge}$ and the partition function with magnetic conductor boundary conditions.
Note that the partition function of a real scalar field has the form $\det{}'(- \nabla^2)^{-1/2}$, so that \eqref{Zedge2} has the form of a wrong-sign scalar field confined to the entangling surface.
This agrees with the divergence found by Kabat \cite{Kabat1995}, as we will show in the next section.

\paragraph{Contact term}

We now show how the edge mode contribution \eqref{Zedge} is related to the contact term.
We will show that the full partition function for the brick wall model (including the edge modes) agrees with the full partition function on the conical manifold (including the curvature coupling).
To do this we regularize the conical geometry in the normal directions by smoothing out the conical singularity \cite{Nelson1994,Fursaev1995,Frolov1995}.
We call the resulting smooth geometry $\B$.

We then Kaluza-Klein reduce Maxwell theory onto $\B$, obtaining towers of massive scalar and vector fields on $\B$, a linear $\sigma$-model of massless scalars whose target space is the space of flat connections on $\mathcal{T}$, and sums over constant electric and magnetic fields.
Details of this reduction will be presented in a companion article \cite{Donnelly:2015hxa}, which will generalize the results of this article to curved spacetimes of the general form $^{(2)} \mathcal{B}\, \times\,^{(D-2)}\!\mathcal{F}$.

The nontrivial curvature coupling comes from the tower of vector fields.
We can calculate this coupling using on-shell duality between massive vector and scalar fields, whose partition functions agree up to a factor proportional to $m^{\chi(\B)}$ coming from the different number of zero modes.
To apply this result to flat spacetime, we take the limit of a disk ($\chi = 1$) with radius $r \to \infty$.  Any extra contributions coming from the boundary at infinity are proportional to $\beta$ and therefore do not affect the geometric entropy \eqref{conical}.  We define the contact term as the product of these extra factors for all modes of $\mathcal{T}$: 
\begin{equation} \label{contact}
Z_\text{contact} = \det{}' \left( \frac{2 \pi \mathrm{Area}(\mathcal{T})}{q^2} (- \nabla^2_\mathcal{T}) \right)^{1/2}
\end{equation}
where a prefactor arising from the massless vector field has been moved into the determinant by rescaling. (The treatment of the boundary at $r = \infty$ is more subtle in the massless case, but in the end it makes no difference \cite{Donnelly:2015hxa}.)

The partition function \eqref{contact} is almost the same as that of the edge modes \eqref{Zedge2}.
The difference comes from the imposition of magnetic conductor boundary conditions, which disallow electric flux through the entangling surface.
The brick wall also changes the Euler characteristic to $\chi(\B) = 0$, and so eliminates the contact term.
The massless scalars, as well as the massive scalars that arise from Kaluza-Klein reduction are assigned Neumann boundary conditions, but the scalar fields dual to massive vectors are Dirichlet.
In the limit $\epsilon \to 0$ introduction of a Neumann boundary changes the partition function only by a local counterterm, relative to no brick wall. 
For the scalar field with Dirichlet boundary conditions there is a difference, which we can calculate as follows.

Consider radial evolution outward from the brick wall in the coordinate $e^r$.
Modes with nonzero angular momentum around the disk decay rapidly away from the entangling surface so their contribution is purely local.
The zero angular momentum modes can be described as a diffusing particle under the evolution $e^{-\frac{1}{2} \alpha p^2}$, where $\alpha = \ln(\epsilon^{-1})/\beta$.
For Neumann boundary conditions, the wave function is initially a $p=0$ eigenstate $\psi_N(p) = \delta(p)$, and so it is invariant under the radial evolution.
The Dirichlet wave function is initially an $x=0$ eigenstate $\psi_D(p) = (2 \pi)^{-1/2}$, and under radial evolution evolves to 
\begin{equation}
(e^{-\frac12 \alpha p^2}\psi_D)(p) = (2 \pi)^{-1/2} e^{-\tfrac12 \alpha p^2} \underset{\epsilon \to 0}{\longrightarrow} \alpha^{-1/2} \delta(p).
\end{equation}
After radial evolution, the Dirichlet wave function approaches the Neumann one up to a constant, so imposing Dirichlet boundary conditions leads to a factor of $\alpha^{-1/2}$ for each mode of the torus ${\cal T}$.

We find that the partition function on a smoothed conical manifold differs from that on a brick wall manifold by the product of the contact term \eqref{contact} and a factor of $\alpha^{1/2}$ for each mode.
This is precisely the edge mode contribution calculated in \eqref{Zedge2}.
Thus we conclude that the effect of coupling to the curvature is to include the contribution of the edge modes to the entropy.

\paragraph{Conformal anomaly} \label{section:anomaly}

To see how the contact term changes universal parts of the entanglement entropy, we consider the logarithmic term in $D=4$, in which Maxwell theory is conformal.
This divergence is universal and related to the conformal anomaly \cite{Solodukhin2008,Casini2011}.
Since the entangling surface is two-dimensional, the contact term will also have a logarithmic divergence.
We now show how this resolves a discrepancy between the entanglement entropy and the conformal anomaly \cite{Dowker2010,Eling2013}.

Letting $\sim$ denote agreement of logarithmic terms, the conformal anomaly predicts that the entanglement entropy of a sphere of radius $r$ is given by \cite{Solodukhin2008,Casini2011}
\begin{equation} \label{Sanom}
S_\text{anom} \sim - \tfrac{31}{45} \ln (r).
\end{equation}
However by a thermodynamic calculation Dowker found \cite{Dowker2010}
\begin{equation} \label{Stherm}
S_\text{therm} \sim -\tfrac{16}{45} \ln (r).
\end{equation} 
We hypothesize that this discrepancy comes from the omission of edge modes from \eqref{Stherm}.

Allowing for electric flux, we find a result proportional to $\det {}' (\Delta_0^E)^{+1/2}$, a negative scalar on the entangling surface $E$ which here is a 2-sphere.
The scalar has a well-known logarithmic divergence, which leads to the same logarithmic divergence in the entropy
\begin{equation}
S_\text{edge} \sim \ln Z \sim - \tfrac{1}{3} \ln (r).
\end{equation}
Thus when the entropy of the edge modes is added to that of the local degrees of freedom \eqref{Stherm}, we find the expected agreement with the conformal anomaly \eqref{Sanom}.

\paragraph{Discussion}

We have shown that Kabat's contact term can be interpreted as a statistical entropy: it is the entanglement entropy of edge modes, i.e. the electric flux $E_\perp$ through the entangling surface.
This gives further support for the thesis of Ref.~\cite{Donnelly2014} that agreement with Euclidean calculations for the entropy requires the inclusion of edge modes.  
The geometric entropy formula \eqref{conical} is also used to calculate the black hole entropy \cite{Gibbons1977}.
Since the entanglement entropy is the same as in flat spacetime to leading order, this resolves a seeming inconsistency with interpreting black hole entropy as entanglement entropy.

Because $E_\perp$ commutes with all gauge-invariant degrees of freedom in the region outside the horizon, it has a continuous configuration space rather than a discrete spectrum; only the constant mode is quantized.
The entropy of this configuration space is the log of its volume, and requires a path integral measure $\D E_\perp$ in \eqref{entropy} to be defined.

If we insist on counting states instead of integrating them, there are actually an infinite number of states associated with even a single nonconstant mode of $E_\perp$. This is a new type of UV divergence, besides that coming from the sum over high frequency modes. Both divergences are regulated by the lattice theory, in which the entangling surface is replaced with finitely many points and $E_\perp$ is quantized on each point.  

Since the entanglement entropy is believed to be finite in a UV-complete theory of quantum gravity \cite{Sorkin1983,Frolov1993,Susskind1994,Jacobson1994,Frolov1996,Jacobson2012,Bianchi2012,Myers2013} presumably this theory must behave like the lattice in the sense of allowing only finitely many states for any mode of $E_\perp$.  
It would be interesting to calculate how accurately one can measure $E_\perp$ before quantum gravity effects become important.

A comment is warranted about the negative sign appearing in the contact term, as this leads to a formally negative expression for the entropy despite its origin as a manifestly positive lattice expression.
Unlike the discrete case, the entropy of a continuous distribution can be negative, and depends on a choice of measure.

The heat kernel regularized partition function of a scalar field is
\begin{equation} \label{heatkernel}
\ln Z = \frac12 \int_{\epsilon^2}^\infty \frac{\tr e^{-s \Delta}}{s} ds = \cdots -\tfrac12 \ln \det \Delta,
\end{equation}
where $\epsilon$ is a UV cutoff length and ``$\cdots$'' are power-law divergences \cite{Hawking1976} (including a constant in even dimensions).
The heat kernel regularization corresponds to a particular choice for these power-law divergences, which determines the leading divergences in the entropy.
In particular, the contribution of a single mode to the heat kernel regularized $\ln Z$ (and hence to the entropy) approaches zero in the UV.
In a scalar field theory, the entropy per mode decreases toward the UV, so that the entropy is positive for each mode below the cutoff.
But for the edge modes, the entropy increases toward the UV (cf. \eqref{Zedge2}), so the heat kernel assigns each mode a negative entropy.
For modes far below the cutoff, this corresponds to the choice of measure 
\begin{equation}
\D E_\perp = \prod_{n > 0} \frac{\epsilon}{\sqrt{2 \pi}} dE_n.
\end{equation}
Had we instead chosen $\D E_\perp$ independent of $\epsilon$ and imposed a hard momentum cutoff, the leading order divergence would have taken a positive rather than a negative sign.
This illustrates graphically the nonuniversality of power law divergences.

Edge modes also appear in nonabelian lattice Yang-Mills theory \cite{Buividovich2008a,Donnelly2011}, so it would be interesting to take a similar continuum limit for that theory.  
It would also be useful to clarify how the entanglement entropy transforms under $p$-form duality (cf. \cite{Agon2013}).
A similar negative contact term also appears in the entropy of gravitons \cite{Fursaev1996,Solodukhin2011}, so it is natural to ask whether this is related to the gravitational edge states.

The negative sign of the heat-kernel-regulated contact term is related to the fact that gauge fields (in low dimensions) and gravitons antiscreen Newton's constant $G$, suggesting a UV fixed point at positive $G$: the asymptotic safety scenario
\cite{Weinberg1979,Niedermaier2006}. 
However, in a scheme where the entropy is inherently positive, such as the lattice regulator, this fixed point will instead occur at negative values of $G$. This might spell trouble for the asymptotic safety program, but we leave this question to future work.

\paragraph*{Acknowledgements}

We are grateful for conversations with Ted Jacobson, Don Marolf, Mark Srednicki, Dan Kabat, Sergey Solodukhin, Joe Polchinski, Chris Eling, Ben Michel, Ed Witten, Josh Cooperman, Markus Luty, Xi Dong, Juan Maldacena, Debajyoti Sarkar, Horacio Casini and Ariel Zhitnitsky.
We also acknowledge the hospitality of Perimeter Institute while part of this work was being completed.
W.D. is supported by funds from the University of California.
A.W. is supported by the Institute for Advanced Study, the Simons Foundation, and NSF Grant No. PHY-1205500.

\bibliographystyle{utphys}
\bibliography{kk-short}

\end{document}